 \documentclass[aip, reprint, twocolumn, amsmath,amsfonts,amssymb]{revtex4-1}

\usepackage{graphicx}
\usepackage{amsmath,amsfonts,amssymb}
\usepackage{color}
\usepackage{bm}
\usepackage{threeparttable}
\usepackage[mathscr]{eucal}

\usepackage{picture}

\newcommand{\bra}{\langle}
\newcommand{\ket}{\rangle}
\newcommand{\bre}{\nonumber\\}

\begin{document}
\title{Second-order perturbation theory with spin-symmetry projected Hartree-Fock}
\author{Takashi Tsuchimochi}
\email{tsuchimochi@gmail.com} 
\affiliation{Graduate School of System Informatics, Kobe University, 1-1 Rokkodai-cho, Nada-ku, Kobe, Japan}
\author{Seiichiro L. Ten-no$^{\dag,}$}
\affiliation{Graduate School of Science, Technology, and Innovation, Kobe University, 1-1 Rokkodai-cho, Nada-ku, Kobe, Japan}
 
\begin{abstract}
We propose two different schemes for second-order perturbation theory with spin-projected Hartree-Fock. Both schemes employ the same {\it ansatz} for the first-order wave function, which is a linear combination of spin-projected configurations. The first scheme is based on the normal-ordered projected Hamiltonian, which is partitioned into the Fock-like component and the remaining two-particle-like contribution. In the second scheme, the generalized Fock operator is used to construct a spin-free zeroth-order Hamiltonian. To avoid the intruder state problem, we adopt the level-shift techniques frequently used in other multi-reference perturbation theories. We describe both real and imaginary shift schemes and compare their performances on small systems. Our results clearly demonstrate the superiority of the second perturbation scheme with an imaginary shift over other proposed approaches in various aspects, giving accurate potential energy curves, spectroscopic constants, and singlet-triplet splitting energies. We also apply these methods to the calculation of spin gaps of transition metal complexes as well as the potential energy curve of the chromium dimer.
\end{abstract}
\maketitle
 
\section{Introduction}
In electronic structure theory, the Schr\"odinger equation is almost always unsolvable because of the exponential growth of the Hilbert space with system size; therefore, the equation is frequently approximated by a computationally solvable model. Such an approach has turned out to be fairly effective for computing many chemically important properties if the model used is well suited to the problem. In most cases, a single determinantal wave function of Hartree-Fock (HF) represents a qualitatively correct wave function at zeroth-order and is employed as a starting point to add the remaining {\it dynamical correlation} effects by accounting for a large number of single and double electron substitutions (SD), each with a small contribution. There are many such single-reference (SR) methods, including M{\o}ller-Plesset perturbation theory (MP),\cite{Moller34} configuration interaction (CI), and coupled-cluster (CC)\cite{Bartlett1981,Bartlett07}. However, there are certain systems where multiple determinants have significant weights in the exact wave function. As a result, HF can introduce tremendous error by neglecting {\it static correlation}, which is a different type of electron correlation than dynamical correlation. To capture static correlation, one has to  consider a multi-configuration (MC) wave function, and significant effort has been made to develop multi-reference (MR) methods that can treat both dynamical and static correlation effects simultaneously.   

The recent advancements and developments in MR methods have been largely based on complete-active-space self-consistent-field (CASSCF).  Arguably, one of the most prominent approaches is CASPT2, i.e., second-order perturbation theory (PT2) with a CASSCF wave function. \cite{Andersson90,Andersson92} CASPT2 has been applied extensively to various applications due to its relatively low computational cost compared to MRCI\cite{Werner88, Knowles88} and MRCC.\cite{Kohn12,Lyakh12} Still, CASPT2 requires the construction of a CASSCF wave function and the diagonalization of a three-particle reduced density matrix (3RDM) within the active space, both of which can often become computational bottlenecks with a large active space.\cite{Kurashige11}


There are other paths to obtain MC wave functions, and one possibility is symmetry-projected HF (PHF).\cite{Jimenez12} It has been well known for a relatively long time that a broken-symmetry determinant $|\Phi_0\ket$ effectively contains multiple determinants as a mixture of states with different symmetries. Among several symmetries, spin-symmetry is considered the most essential symmetry that HF violates in order to introduce static correlation. Hence, applying a spin-projection operator $\hat P$ to unrestricted HF (UHF) makes it possible to generate a compact MC wave function $\hat P|\Phi_0\ket$. In practice, molecular orbitals in $|\Phi_0\ket$ are relaxed self-consistently in the presence of $\hat P$ by minimizing its energy, and as a result, $\hat P|\Phi_0\ket$ can be regarded as a relatively efficient and reasonable MCSCF wave function. This method is referred to as spin-projected UHF (SUHF) and is expected to offer a suitable platform for subsequent dynamical correlation treatment. 

It should be noted that the concept of spin-projection emerged in the seminal work of L\"owdin in the mid-1950s.\cite{Lowdin55B} However, the difficulty of handling the many-body nature of a spin-projection operator has long hindered the development of its extension to treating dynamical correlation.\cite{Mayer73,Mayer80} The first post-PHF method was proposed by Schlegel in 1986,\cite{Schlegel86,Schlegel88} followed shortly after by Knowles and Handy,\cite{Knowles88A,Knowles88B} where spin-unrestricted MP2 (UMP2) was approximately spin-projected. Only recently has spin-extended MP2 (EMP2) been introduced, which performs numerically {\it exact} spin-projection onto an MP1 wave function constructed from the underlying broken-symmetry determinant $|\Phi_0\ket$ of SUHF (rather than UHF).\cite{Tsuchimochi14} Since then, various post-SUHF methods have been developed, including time-dependent SUHF,\cite{Tsuchimochi15A} CI, \cite{Tsuchimochi16A,Tsuchimochi16B} and CC.\cite{Duguet15,Tsuchimochi17B,Qiu17B,Qiu18,Tsuchimochi18,Tsuchimochi19A} These methods have been shown to generally outperform their restricted and unrestricted variants, especially when static correlation plays a key role. However, in the course of numerous test applications of the developed methods, we have found that the improvements the original EMP2 has to offer are somewhat limited, given the considerable improvements of spin-projected CI over unrestricted CI.\cite{Tsuchimochi16A,Tsuchimochi16B} For instance, while the original EMP2 works well for biradicals, such as single-bond dissociation, its accuracy becomes substantially worse for more complicated cases, such as double and triple bond breaking, as will be discussed below. Furthermore, the predetermined nature of the first-order wave function does not allow the corresponding Hylleraas functional to be defined,\cite{Sinanoglu61,Hylleraas30} which would be useful in developing the geometry optimization method.\cite{Celani03,Park19} Given that perturbation theory is not unique and its performance is greatly dependent on the choice of zeroth-order Hamiltonian, we believe it is desirable to continue exploring the possibility of more appropriate perturbation schemes for SUHF.
 
To this end, in this paper, we propose and test two perturbative corrections on SUHF. The first one is regarded as a generalization of the original EMP2 of Tsuchimochi and Van Voorhis,\cite{Tsuchimochi14} which will be referred to as EMP2(0) hereafter to distinguish it from the newly developed EMP2 in the present work. It is based on the normal-ordered Hamiltonian introduced for the nonorthogonal determinants that appear in the integration of spin-projection.\cite{Tsuchimochi17B}  In the second scheme, which we call SUPT2, the so-called generalized Fock matrix is used as a starting point, as in CASPT2.\cite{Andersson90,Andersson92} Consequently, SUPT2 shares many common properties as well as limitations with CASPT2. Indeed, it will be demonstrated below that the notorious intruder state problem is also inevitable in SUPT2, and we therefore also develop the level-shift technique frequently used in CASPT2.\cite{Roos95,Forsberg97,Ghigo04} 
In this work, their performances are compared by using simple test systems as well as transition metal complexes. 

This paper is organized as follows. Section \ref{sec:SUHF} presents an overview of SUHF. In Section \ref{sec:RSPT2}, we apply the Rayleigh-Schr\"odinger perturbation theory with an SUHF reference, and consider two possible {\it ans\"atze} for the first-order wave function. Section \ref{sec:EMP2} reviews EMP2(0) and proposes the generalized EMP2, while Section \ref{sec:SUPT2} describes the SUPT2 theory. We introduce real and imaginary level-shifts in Section \ref{sec:Level-shift}, the latter of which requires some elaboration. 
Section \ref{sec:Results} first presents a comparison between several methods tested for the HF, H$_2$O, and N$_2$ molecules and discusses the intruder state problem in SUPT2. It also presents the results for the spectroscopic constants of N$_2$, singlet-triplet splitting energies of various systems including transition metal complexes, and the potential energy curve of the Cr$_2$ molecule.  In Section \ref{sec:Discussions}, we discuss the main cause of the different behaviors between EMP2 and SUPT2.  Finally, conclusions are drawn in Section \ref{sec:Conclusions}.

\section{Theory}
\subsection{Spin-projected unrestricted Hartree-Fock}\label{sec:SUHF}
Here, we briefly review SUHF and define some quantities that will be required in the following sections. Below, $i,j,k,$ and $l$ will represent occupied spin-orbitals in $|\Phi_0\ket$, and $a,b,c,$ and $d$ represent virtual spin-orbitals. General spin-orbitals are denoted by $p,q,r,$ and $s$. Because our approach is based on spin-unrestricted orbitals, in some cases, we will use $\sigma=\alpha,\beta$ to specify the spin of orbitals. Capital letters are used for spin-restricted orbitals. 

In this work, a spin-projection operator $\hat P$ is given by the following form:
\begin{align}
	\hat P =  \frac{2S+1}{8\pi^2}\int_\Omega d\Omega\; w(\Omega) \hat R(\Omega), 
\end{align}
 where $\Omega=(\alpha,\beta,\gamma)$ are the Euler angles, $w(\alpha,\beta,\gamma)$ Wigner's D-matrix elements representing {\it fixed} weights, and 
\begin{align}
	\hat R(\alpha,\beta,\gamma) = e^{-i \alpha \hat S_z}e^{-i \beta \hat S_y}e^{-i \gamma \hat S_z}
\end{align}
 the spin-rotation operators. 
 Accordingly, $\hat R(\Omega)|\Phi_0\ket$ gives a different determinant that is not orthogonal to $|\Phi_0\ket$. Discretizing $\hat P$ with $N_g$ grid points labeled by $g$, we write an SUHF wave function as
\begin{align}
	\hat P|\Phi_0\ket = \sum_g^{N_g} w_g \hat R_g |\Phi_0\ket,
\end{align}
which is regarded as a linear combination of nonorthogonal determinants. Because $\hat P$ is idempotent, Hermitian, and commutable with the non-relativistic Hamiltonian $\hat H$, the SUHF energy is simply given by
\begin{align}
	E_{\rm SUHF} = \frac{\bra \Phi_0|\hat H \hat P|\Phi_0\ket }{\bra \Phi_0|\hat P|\Phi_0\ket}.
\end{align}
The variational principle applied to SUHF gives the generalized Brillouin theorem:
\begin{align}
    \bra \Phi_0|\hat a_a^{i}(\hat H - E_{\rm SUHF})\hat P|\Phi_0\ket  =0, \label{eq:BT}
\end{align}
where $\hat a_q^p$ are single excitation operators from the $q$th to $p$th orbital.

It will prove useful later to introduce the normal-ordered products $\{\cdots\}_g$ for two nonorthogonal determinants $|\Phi_0\ket$ and $\hat R_g|\Phi_0\ket$,\cite{Tsuchimochi16A,Tsuchimochi16B} meaning
\begin{align}
	\bra \Phi_0|\{\cdots\}_g \hat R_g|\Phi_0\ket \equiv 0.
\end{align}
Using this definition, it is easy to show that the second-quantized Hamiltonian $\hat H$ can be written as\cite{Tsuchimochi17B}
\begin{align}
	\hat H &= \sum_{pq} h_{pq} \hat a^p_q + \frac{1}{4}\sum_{pqrs}\bra pq||rs\ket \hat a^{pq}_{rs}\\
	&=E_g + \sum_{pq} ({\bf F}_g)_{pq} \{\hat a^{p}_{q}\}_g + \frac{1}{4}\sum_{pqrs}\bra pq||rs\ket \{\hat a^{pq}_{rs}\}_g,\label{eq:Normal}
\end{align}
for {\it any} $g$, where $\bra pq||rs\ket$ are the standard anti-symmetrized two-electron integrals, and
\begin{align}
&	E_g = \frac{\bra \Phi_0|\hat H \hat R_g|\Phi_0\ket}{\bra \Phi_0|\hat R_g|\Phi_0\ket}\\
&	({\bf F}_g)_{pq} = h_{pq} + \sum_{rs} \bra pr||qs\ket \frac{\bra \Phi_0|\hat a^r_s\hat R_g|\Phi_0\ket}{\bra \Phi_0|\hat R_g|\Phi_0\ket},
\end{align}
are the transition energy and transition Fock matrix, respectively. The required matrix elements in this work can be easily derived using the Wick theorem extended to the nonorthogonal representation.\cite{Tsuchimochi16A} For further details, the reader can refer to Refs.~[\onlinecite{Jimenez12,Tsuchimochi16A, Tsuchimochi16B}].

\subsection{Perturbation Theory}\label{sec:RSPT2}
  In the Rayleigh-Schr\"odinger perturbation theory, the Hamiltonian is partitioned as
  \begin{align}
  	\hat H = \hat H_0 + \lambda \hat V,
  \end{align}
  and the exact FCI wave function and its energy are expanded as
\begin{align}
&|\Psi\ket = |\psi_0\ket + \lambda |\psi_1\ket + \lambda^2 |\psi_2\ket + \cdots,\\
& E = E_0 + \lambda E_1 + \lambda^2 E_2 + \cdots. \label{eq:FCI}
\end{align}
The choice of $\hat H_0$ is left arbitrary, and will thus be determined later. As is well known, the order-by-order expansion of the Schr\"odinger equation results in
\begin{align}
	&\hat H_0 |\psi_0\ket = E_0 |\psi_0\ket,\label{eq:0th}
	\end{align}
	and
	\begin{align}
	&\left(\hat H_0 -E_0\right) |\psi_n\ket + \hat V|\psi_{n-1}\ket = E_n |\psi_0\ket + \sum_{k=1}^{n-1} E_{n-k} |\psi_k\ket.\label{eq:1st}\end{align}

 In this work, we wish to formulate a perturbation theory using an SUHF wave function as the reference zeroth-order wave function:
\begin{align}
	|\psi_0\ket \equiv \hat P |\Phi_0\ket.
\end{align}
To do so, first we have to develop an {\it ansatz} for $|\psi_1\ket$ for the second-order energy $E_2$. Generally, higher order wave functions have to be cleanly separated from the reference state. This means they are orthogonal to each other:
\begin{align}
	\bra \psi_0|\psi_1\ket = 0.
\end{align}
This can be accomplished by defining the projection operator that projects onto the reference space
\begin{align}
\hat {\cal P}_0 \equiv \frac{|\psi_0\ket \bra \psi_0|}{\bra \psi_0|\psi_0\ket} =\frac{\hat P|\Phi_0\ket \bra \Phi_0|\hat P}{\bra \Phi_0 | \hat P |\Phi_0\ket}
\end{align}
and its complementary projector
\begin{align}
\hat {\cal Q}_0 = 1 - \hat {\cal P}_0.
\end{align} 
Using $\hat {\cal Q}_0$, $|\psi_1\ket$ can be generally expanded as 
\begin{align}
	|\psi_1\ket = \sum_{\Omega} \hat {\cal Q}_0|\Omega\ket t_\Omega,
\end{align}
where the basis $\{|\Omega\ket\}$ spans the first-order interacting space of $|\psi_0\ket$, and $t_\Omega$ are the amplitude coefficients. The form of $\{|\Omega\ket\}$ needs to be determined.

As in standard MRPT2 schemes, a natural choice for $\{|\Omega\ket\}$ would be internally-contracted configurations with respect to an SUHF wave function. In this case, only the singles and doubles spaces are needed, although the former does not contribute to the second-order energy if the Brillouin theorem is satisfied. Therefore, the unitary-group-generator $\hat {\cal E}_{\Omega}$ may be used to produce such a basis:
\begin{align}
	|\Omega\ket = \hat {\cal E}_\Omega \hat P |\Phi_0\ket. \label{eq:ic}
\end{align}
Viewing SUHF as a type of MCSCF, it has an incomplete active space, where $N_e$ electrons are correlated in $N_e$ active orbitals, while there is an intrinsic secondary space whose occupations are strictly zero.\cite{Tsuchimochi16B} Thus, there are less double excitation sub-blocks to be considered than in other MRPT2 schemes, and they can be categorized as one of the following sub-blocks: fully-internal, semi-external, and external excitations, where zero, one, and two electrons are excited to the virtual space, respectively. The fully-internal excitations are those within the active space, and they are neglected in CASPT2\cite{Andersson90,Andersson92} under the assumption that a CAS does not change in the presence of dynamical correlation. This type of excitation is also missing in other MRPT2 theories that use an incomplete model space\cite{Malmqvist08, Ma16, Kahler17} because it would give rise to significant complication or a large number of intruder states. The exclusion of fully-internal excitations may be valid if the incomplete active space is almost complete. However, this is far from the case for SUHF. Therefore, one must consider excitations into almost fully occupied orbitals or from nearly empty ones, introducing significant redundancies. Given this fact, this ``excitation-after-projection'' scheme, as given in Eq.~(\ref{eq:ic}), is not advantageous as it is likely to bring significant complication to the derivation, while most fully-internal excitations are redundant. 

The above difficulty can be avoided by exploiting the compact representation of the SUHF wave function. Namely, in the ``projection-after-excitation'' {\it ansatz}, we write
\begin{align}
|\Omega \ket =  \hat P \hat a_\Omega|\Phi_0\ket,
\end{align}
where broken-symmetry excitation operators $\hat a_\Omega$ generate a series of excited determinants with respect to $|\Phi_0\ket$, such as $|\Phi_i^a\ket$ and $|\Phi_{ij}^{ab}\ket$, which are then projected by $\hat P$. Because there is a clear distinction between occupied and virtual orbitals in $|\Phi_0\ket$, all $|\Omega\ket$ are realistic with a large norm. Nevertheless, we should note that the projection-after-excitation basis is still slightly redundant due to the nature of $\hat P$, which includes not only excitations but also de-excitations.\cite{Tsuchimochi18}

Using the shorthand $|\Phi_\mu\ket = \hat a_\mu |\Phi_0\ket$, we write the first-order wave function as
\begin{align}
	|\psi_1\ket = \sum_{\mu} \hat {\cal Q}_0\hat P  |\Phi_\mu\ket t_\mu. \label{eq:PT1}
\end{align}
It should be stressed that in the above equation, only projected singles and doubles are essential for expanding $|\psi_1\ket$. The projected-excited determinants of higher-rank could be included in $|\psi_1\ket$ because they in fact interact with $\hat P|\Phi_0\ket$ through $\hat H$. However, it is expected that their contributions should be negligible or even non-existent, as it can be easily shown that $\{\hat P|\Phi_i^a\ket, \hat P|\Phi_{ij}^{ab}\ket\}$ span exactly the first-order interacting space with respect to $\hat P|\Phi_0\ket$.\cite{Tsuchimochi16B} It is also noteworthy that the projected singles and doubles include the space corresponding to the fully-internal excitations of the excitation-after-projection scheme and are thus potentially capable of relaxing the SUHF (incomplete) active space.

Using Eq.~(\ref{eq:PT1}), the second-order energy can be given by 
\begin{align}
E_2 &=  \bra \psi_0 | \hat H |\psi_1 \ket =\sum_\mu^{\rm SD} \bra  \Phi_0|\hat P \hat H \hat {\cal Q}_0 \hat P |\Phi_\mu\ket t_\mu,\label{eq:E2}
\end{align}
which, noting that there is no contribution from singles due to the Brillouin theorem Eq.~(\ref{eq:BT}), becomes
\begin{align}
E_2& =  \sum_{i>j}\sum_{a>b}  \bra \Phi_0 | \left(\hat H - E_{\rm SUHF}\right) \hat P|\Phi_{ij}^{ab} \ket t_{ij}^{ab}.
\end{align}
It should be pointed out that this expression is identical to that of the second-order energy of EMP2(0).\cite{Tsuchimochi14} 
 The amplitudes {\bf t} are determined by projecting the first-order Eq.~(\ref{eq:1st}) with the manifold $\{\hat {\cal Q}_0 \hat P |\Phi_\mu\ket\}$:
\begin{align}
	\sum_\nu^{\rm SD} \bra \Phi_\mu | \hat P \hat {\cal Q}_0 \left(\hat H_0 - E_0\right) \hat {\cal Q}_0\hat P|\Phi_\nu\ket t_\nu + \bra  \Phi_\mu|\hat P\hat {\cal Q}_0 \hat H  \hat P |\Phi_0\ket = 0,\label{eq:amp}
	\end{align}
which can be simplified to
\begin{align}
	\sum_\nu^{\rm SD} A_{\mu\nu} t_\nu + v_\mu = 0,\label{eq:At=v}
\end{align}
with
\begin{align}
	A_{\mu\nu} &= \bra \Phi_\mu | \hat P \hat {\cal Q}_0 \left(\hat H_0 - E_0\right) \hat {\cal Q}_0\hat P|\Phi_\nu\ket,\\
	v_\mu &= \bra \Phi_\mu|\left(\hat H - E_{\rm SUHF}\right)\hat P |\Phi_0\ket.\label{eq:Amat}
\end{align}
Thus, the linear equation depends on the choice of zeroth-order Hamiltonian $\hat H_0$. It is noteworthy that Eq.~(\ref{eq:At=v}) resembles the amplitude equations of other MR methods. In these methods, the matrix that corresponds to {\bf A} is often diagonalized in each excitation sub-block, which is feasible if 3RDM can be diagonalized.\cite{Andersson92,Malmqvist08,Ma16}  The linear dependence is also removed through this procedure.\cite{Andersson90} On the contrary, in our projection-after-excitation scheme, there appears to be no such separable sub-blocks of excitations, and therefore {\bf A} cannot be diagonalized. However, {\bf A} is generally sparse regardless of the choice of $\hat H_0$\cite{Tsuchimochi16B} if the orbital set used is biorthogonal between $\alpha$ and $\beta$ spins.\cite{Amos61} Also, the linear dependence in {\bf A} shows up in {\bf v} in exactly the same manner,\cite{Tsuchimochi18,footnote_SUPT2_2}  so it need not be removed in practice. 
Thus, the linear Eq.~(\ref{eq:At=v}) can be directly solved.

We note that singles should be explicitly treated when solving Eq.~(\ref{eq:At=v}). Otherwise, convergence is usually not obtained. This is because the projected singles and doubles are not orthogonal to each other (due to the redundancy in our scheme),  and the linear dependence would not be treated correctly if without singles. In any case, the singles space is trivial in size and is required when the generalized Brillouin theorem is not satisfied, which is often the case in our illustrative calculations below. Therefore, we always include single excitations throughout this work.  

It is well known that perturbation theory can be formulated as a variational problem.\cite{Sinanoglu61, Hylleraas30} Namely, one can define the Hylleraas functional
\begin{align}
	{\cal L} = \bra \psi_1 | \left(\hat H_0 - E_0 \right) |\psi_1\ket + 2\bra \psi_1| \hat H |\psi_0\ket,\label{eq:Hylleraas}
\end{align}
whose stationary point corresponds to the second-order energy $E_2$. Eq.~(\ref{eq:amp}) appears as a consequence of the variational principle of ${\cal L}$ with respect to the amplitudes. With ${\cal L}$, it is rather straightforward to adopt the standard derivative methods.\cite{Jorgensen88, Helgaker89}

Now that we have established a general perturbation theory with SUHF based on the projection-after-excitation scheme, only a definition of $\hat H_0$ is now required, which is somewhat arbitrary. Nevertheless, it is widely known that the choice of $\hat H_0$ significantly affects the final performance, and it should therefore be carefully chosen. To end this section, we remark on a few preferable conditions that $\hat H_0$ should hold:
\begin{enumerate}
	\item It must have $|\psi_0\ket = \hat P|\Phi_0\ket$ as its eigenstate. 
 In this work, we employ a spin-free zeroth-order Hamiltonian so that $[\hat H_0, \hat P] = 0$, which allows for a considerable simplification, although this is by no means a requisite condition.
	\item It should be chosen such that the perturbation $\hat V$ is sufficiently small.
	\item It should be composed of one-electron operators for ease of derivation and computation.
	\item It should reduce to the standard Fock operator in the absence of $\hat P$ so as to reproduce the MP$n$ energies.
\end{enumerate}
In the following sections, we will consider two possibilities for the form of $\hat H_0$ based on these guidelines.

\subsection{EMP2}\label{sec:EMP2}
The original EMP2(0) also starts with the same {\it ansatz} for $|\psi_1\ket$, i.e., Eq.~(\ref{eq:PT1}).\cite{Tsuchimochi14} Without explicitly defining $\hat H_0$, its first-order wave function is fixed to the spin-projected MP1 wave function. The amplitudes are obtained by semi-canonicalization of spin-contaminated UHF-like Fock matrices, where one separately diagonalizes the occupied-occupied and virtual-virtual blocks of the spin-dependent Fock matrices computed with broken-symmetry $|\Phi_0\ket$.\cite{Lauderdale91}
This circumvents iterative calculations when solving Eq.~(\ref{eq:amp}), which are otherwise necessary because $\hat H_0$ is generally not diagonal in the working basis $\{\hat {\cal Q}_0 \hat P |\Phi_\mu\ket\}$. While EMP2(0) does go back to standard MP2 when $\hat P$ is neglected, it remains largely unclear with respect to what energy {\bf t} is optimized; hence, the derivation of analytical derivatives would become complicated.

A somewhat more general formalism can be derived by using the normal-ordered Hamiltonian in Eq.~(\ref{eq:Normal}). The idea is to write the projected Hamiltonian $\hat H\hat P$ as
\begin{align}
	\hat H \hat P = \hat {\cal H}_0 + \hat {\cal V},
\end{align}
with
\begin{align}
	&\hat {\cal H}_0 = \sum_g^{N_g} w_g \left(E_g \hat R_g + \sum_{pq}({\bf F}_g)_{pq} \{\hat a^p_q\}_g \hat R_g\right),\\
	&\hat {\cal V}  = \sum_g^{N_g} w_g \left( \frac{1}{4} \sum_{pqrs}\bra pq||rs\ket \{\hat a^{pq}_{rs}\}_g\hat R_g\right).
\end{align}
In our previous study on spin-extended CISD (ECISD),\cite{Tsuchimochi16B} it was found that the contribution of $\hat {\cal V}$ is typically small compared to that of $\hat {\cal H}_0$; thus, the latter was used as preconditioning in the iterative diagonalization of the ECISD Hamiltonian. This indicates that $\hat {\cal H}_0$ is reasonable for a zeroth-order component of the projected Hamiltonian. 
Because $\hat {\cal H}_0$ does not have $\hat P|\Phi\ket$ as its eigenstate in general, one can formally define the following zeroth-order Hamiltonian for EMP2:
\begin{align}
	\hat H_0^{\rm EMP2} = \hat {\cal P}_0 \hat {\cal H}_0 \hat {\cal P}_0 + \hat {\cal Q}_0 \hat {\cal H}_0 \hat {\cal Q}_0.
\end{align}
However, because $\hat {\cal H}_0$ is not spin-free, the matrix elements of $\hat P\hat {\cal H}_0 \hat P$, including the zeroth-order energy
\begin{align}
	E_0^{\rm EMP2} = \frac{\bra \Phi_0| \hat P  \hat {\cal H}_0 \hat P |\Phi_0\ket}{\bra \Phi_0|\hat P |\Phi_0\ket },
\end{align}
become cumbersome to evaluate;  the required number of grid points becomes $N_g^3$, which adds considerable computational overhead.  To alleviate this problem, we simply introduce the following approximation:
\begin{align}
	\bra \Phi_\mu | \hat P \hat {\cal H}_0 \hat P| \Phi_\nu\ket \approx 	\bra \Phi_\mu | \hat {\cal H}_0 | \Phi_\nu\ket.\label{eq:approx}
\end{align}
We deem this approximation to be reasonable as $\hat {\cal H}_0$ itself plays a role of approximate spin-projection. In fact, if $\hat {\cal V}$ is negligible, which is our assumption in EMP2, then $ \hat {\cal H}_0 \approx \hat H \hat P$, and therefore Eq.~(\ref{eq:approx}) certainly holds.  One caveat is that the perturbation series would not converge to the correct limit, Eq.(\ref{eq:FCI}). 

 With Eq. (\ref{eq:approx}), the zeroth-order energy $E_0$ is simply the SUHF energy. By absorbing $E_{\rm SUHF}$ in $\hat {\cal H}_0$ and defining
\begin{align}
	 \hat {\tilde {\mathcal H}}_0   = \sum_g^{N_g} w_g \left( (E_g -E_{\rm SUHF}) \hat R_g + \sum_{pq}({\bf F}_g)_{pq} \{\hat a^p_q\}_g\right),
\end{align}
the amplitude Eq.~(\ref{eq:amp}) becomes
\begin{align}
	\sum_\nu^{\rm SD} \Bigl[&\bra \Phi_\mu| \hat {\tilde {\mathcal H}}_0   |\Phi_\nu\ket  
	- \bra \Phi_\mu|\hat {\tilde {\mathcal H}}_0  |\Phi_0\ket \bra \Phi_0|\hat P |\Phi_\nu\ket\bre
	&- \bra \Phi_\mu|\hat P|\Phi_0\ket \bra \Phi_0|\hat {\tilde {\mathcal H}}_0   |\Phi_\nu\ket\Bigr] t_\nu \bre
	&+ \bra \Phi_\mu | \left(\hat H - E_{\rm SUHF}\right)\hat P|\Phi_0\ket = 0, \label{eq:EMP2eq}
\end{align}
which means the matrix {\bf A} can be expressed as
\begin{align}
	 A^{\rm EMP2}_{\mu \nu} &=\bra \Phi_\mu| \hat {\tilde {\mathcal H}}_0   |\Phi_\nu\ket  
	- \bra \Phi_\mu|\hat {\tilde {\mathcal H}}_0  |\Phi_0\ket \bra \Phi_0|\hat P |\Phi_\nu\ket\bre
	&- \bra \Phi_\mu|\hat P|\Phi_0\ket \bra \Phi_0|\hat {\tilde {\mathcal H}}_0   |\Phi_\nu\ket.
\end{align}
 Incidentally, we note that the EMP2(0) amplitudes can be obtained as a special case by assuming no rotation is done ($\hat R_g =\hat 1$) in the above Eq.~(\ref{eq:EMP2eq}), i.e., no spin-projection is performed. In such a case, one can easily find an orbital basis that diagonalizes the matrix elements in the first term of the equation: the semi-canonical orbital basis. On the other hand, this generalized EMP2 Eq.~(\ref{eq:EMP2eq}) is nonorthogonal and contains off-diagonal elements; thus, it is solved iteratively as described in the previous section. The Hylleraas functional for EMP2 is straightforward to derive using these approximate matrix elements.

\subsection{SUPT2}\label{sec:SUPT2}
While the derivation of EMP2 in the previous section is largely specific to the nonorthogonal structure of $\hat P$, it is also interesting to incorporate and combine the conventional wisdom of established MR perturbation theories. To this end, we will closely follow the approach taken by CASPT2.\cite{Andersson90,Andersson92} This perturbation scheme is therefore called SUPT2, and it is based upon the spin-average generalized Fock operator
\begin{align}
\hat F = \sum_{PQ} f_{PQ} \left(\hat a^{P\alpha}_{Q\alpha} + \hat a^{P\beta}_{Q\beta} \right), \label{eq:Fock}
\end{align} 
where the generalized Fock matrix {\bf f} is given in the same manner as in CASPT2, i.e., through the 1RDM ${\bf D}$ of the reference wave function
\begin{align}
 f_{PQ} = h_{PQ} + \sum_{RS} D_{SR} \Bigl[\bra PR|QS\ket - \frac{1}{2} \bra PR|SQ\ket\Bigr].
\end{align} 
Then, a zeroth-order Hamiltonian may be defined as
\begin{align}
	\hat H_0 = \hat {\cal P}_0 \hat F \hat {\cal P}_0 + \hat {\cal Q}_0 \hat F \hat {\cal Q}_0.
\end{align}
The important point here is that $[\hat F,\hat P] = 0$, which allows for the desired eigenvalue equation,
\begin{align}
	\hat H_0 \hat P| \Phi_0\ket = E_0 \hat P |\Phi_0\ket,
\end{align}
where the zeroth-order energy is 
\begin{align}
	E_0 = \bra \Phi_0|\hat F \hat P |\Phi_0\ket = \sum_{PQ} f_{PQ} D_{QP}.
\end{align}
 
Using $\hat H_0 {\hat {\cal P}}_0\equiv E_0 {\hat {\cal P}}_0$, it is easy to show that the ${\bf A}$ matrix in Eq.~(\ref{eq:Amat}) is 
\begin{align}
A^{\rm SUPT2}_{\mu\nu} &= \left(F_{\mu\nu} -E_0 S_{\mu\nu}\right) \bre
&\;\;\;\;\; - S_{\mu 0} \left(F_{0\nu} - E_0 S_{0\nu}\right) -\left( F_{\mu 0} -E_0 S_{\mu 0}\right) S_{0 \nu}
\end{align}
with the projected matrix elements
\begin{align}
	&F_{\mu\nu} = \bra \Phi_\mu|\hat F \hat P |\Phi_\nu\ket, \\
	&S_{\mu\nu} = \bra \Phi_\mu| \hat P |\Phi_\nu\ket, 
\end{align}
which can be straightforwardly evaluated.

\subsection{SUPT2 with a shift operator} \label{sec:Level-shift}
In our preliminary calculations, it was found that SUPT2 suffers from intruder states. This happens whenever some eigenvalues of $\hat H_0$ in the orthonormal space, in which the overlap metric is diagonal, are nearly degenerate with $E_0$. The so-called intruder state problem is notoriously common in CASPT2, especially if the active space is small, and the de-facto standard to ameliorate this issue is to shift the zeroth-order Hamiltonian by a real constant $\epsilon$:\cite{Roos95}
\begin{align}
\hat H_0 \rightarrow \hat H_0 + \epsilon \hat {\cal Q}_0.\label{eq:real}
\end{align}
A typical choice for $\epsilon$ in CASPT2 is 0.2$\sim$0.3 Hartree ($E_{\rm H}$). It is straightforward to also use the above level-shifted $\hat H_0$ for SUPT2. $E_2$ of the real-shifted SUPT2 (rSUPT2) is underestimated due to the positive shift $\epsilon$, but this is usually corrected by using the Hylleraas functional Eq.~(\ref{eq:Hylleraas}) instead:
\begin{align}
{\cal L}^{\rm rSUPT2} = E_2 - \epsilon \bra\psi_1| \psi_1 \ket.\label{eq:correct}
\end{align}
As will be shown below, such a level shift mitigates the ill-behaved energy profiles of SUPT2. However, a real level shift merely moves the positions of singularities, as the eigenvalues are likely to continuously change between negative and positive values when moving along a potential surface. Therefore, there is always a chance of divergence because the shifted eigenvalues can still be accidentally close to $E_0$. Prior to calculations, one does not know how large $\epsilon$ should be to guarantee that all eigenvalues are above $E_0$. Also, the level-shift corrected energy (Eq.~(\ref{eq:correct})) is not stationary with respect to the amplitudes, and its derivative requires appropriate Lagrange multipliers. One could simply use the uncorrected $E_2$, but it increasingly deteriorates with larger $\epsilon$. 

It is more appealing to use an imaginary level shift $i\epsilon$, which completely removes the singularities at the cost of slight distortion in the potential surface.\cite{Forsberg97} With an imaginary level shift, the poles are shifted towards the imaginary axis and never appear on the real axis, on which one evaluates the energy. Another advantage of the imaginary level shift is that, away from the poles, the energy change induced by $i\epsilon$ is much smaller than that with the real level shift $\epsilon$.\cite{Forsberg97,Park19} However, a disadvantage is that applying an imaginary shift to SUPT2 is not as straightforward, because the original implementation for CASPT2 assumes an orthonormal basis, which is not tractable to compute in SUPT2. Below, we therefore formulate an imaginary level shift scheme in a slightly different way.

Suppose that we have successfully diagonalized the {\bf A} matrix (which we never do in practice) and obtained eigenvalues
\begin{align}
\sum_\nu^{\rm SD} A_{\mu\nu} U_{\nu\tilde \mu} =  \Delta_{\tilde \mu} U_{\mu\tilde \mu} \label{eq:diag}.
\end{align}
Note that in solving Eq.~(\ref{eq:diag}), linearly dependent solutions are discarded. Also, note that exactly the same redundancy is shared by $v_\mu$.
The unitary matrix {\bf U} transforms $\{|\Phi_\mu\ket\}$ to $\{|\tilde \Phi_{\tilde \mu}\ket \}$ with
\begin{align}
	|\tilde \Phi_{\tilde\mu}\ket = \sum_\mu^{\rm SD}|\Phi_\mu\ket U_{\mu \tilde\mu},\label{eq:tildemu}
\end{align}
which thus gives the following diagonal representation: 
\begin{align}
\bra \tilde \Phi_{\tilde\mu}| \hat P \hat {\cal Q}_0 \left(\hat H_0 - E_0 \right)\hat {\cal Q}_0 \hat P | \tilde \Phi_{\tilde \nu}\rangle = \Delta_{\tilde\mu}\delta_{\tilde \mu\tilde \nu}.
\end{align}
 Importantly, the projected basis $\{\hat {\cal Q}_0\hat P|\tilde \Phi_{\tilde \mu}\ket\}$ is {\it not} orthonormal. In other words, we skip the orthogonalization step employed in CASPT2 and directly diagonalize $\hat P\hat {\cal Q}_0\left(\hat H_0 -E_0\right)\hat {\cal Q}_0\hat P$ as a whole.  

The amplitudes $T_{\tilde \mu}$ in this diagonal basis are simply given by 
\begin{align}
	T_{\tilde \mu} = -\frac{V_{\tilde \mu}}{\Delta_{\tilde\mu}},\label{eq:Tmu}
\end{align}
where
\begin{align}
	V_{\tilde\mu} =\bra \tilde \Phi_{\tilde \mu}| \hat P \hat {\cal Q}_0 \hat H   \hat P |\Phi_0\ket = \sum_\mu U^*_{\mu{\tilde \mu}}  v_\mu.
\end{align}
Nearly zero $\Delta_{\tilde \mu}$ (ones not caused by the linear dependency) obviously give rise to a divergence in the amplitudes and thus in the second-order energy. In the proposed imaginary-shifted SUPT2 (iSUPT2), the denominator is directly regularized by $i\epsilon$ instead of changing the zeroth-order Hamiltonian like in Eq.~(\ref{eq:real}):
\begin{align}
	T_{\tilde \mu} &\rightarrow -\frac{\bra \tilde \Phi_{\tilde \mu}| \hat P\hat {\cal Q}_{0}  \hat H  \hat P   |\Phi_0\ket  }{\Delta_{\tilde\mu} + i\epsilon},
	\end{align}
where only the real part is used for evaluation of the second-order energy in order to avoid complex algebra. Namely, our imaginary-shifted amplitudes are {\it defined} as
\begin{align}
	{\mathscr T}_{\tilde \mu} 
	&\equiv -\frac{\bra \tilde \Phi_{\tilde \mu}| \hat P\hat {\cal Q}_{0}  \hat H  \hat P   |\Phi_0\ket \Delta_{\tilde \mu}}{\Delta_{\tilde\mu}^2 + \epsilon^2}, \label{eq:ishiftamp}
\end{align}
which are apparently singularity-free. To obtain the working amplitude equation, we back-transform Eq.~(\ref{eq:ishiftamp}) using Eqs.~(\ref{eq:diag} and \ref{eq:tildemu}) to get
\begin{align}
\sum_{\lambda\nu}A_{\mu\lambda}A_{\lambda \nu} t_\nu + \sum_\nu A_{\mu\nu}v_\nu   + \epsilon^2 t_\mu  	
= 0,\label{eq:A2}
\end{align}
where we have used the unitarity of {\bf U}. The iSUPT2 energy is obtained by substituting the converged {\bf t} into the Hylleraas functional; again, such an energy is not stationary with respect to the amplitudes. 
The equation is quadratic in {\bf A}, but this can be easily handled by forming ${\bf A}{\bf x}$ twice, i.e., ${\bf A}{\bf t}$ followed by ${\bf A}\left({\bf At} + {\bf v}\right)$. Hence, the computational cost is doubled, which is still much better than diagonalizing the entire matrix to compute $\Delta_{\tilde \mu}$ explicitly. 

We should stress that the above approach is different from the use of the modified zeroth-order Hamiltonian $\hat H_0 + i\epsilon \hat {\cal Q}_0$. The former is deemed to be more beneficial because it does not require the diagonalization of the overlap matrix to obtain an orthonormal basis while the latter does. Nonetheless, this difference results in a very minor change in the final energy in our experience.

Lastly, we note that EMP2 is almost always free from the intruder state problem because {\bf A} is thought of as an approximation of the ECISD Hamiltonian, neglecting the two-particle-like operator $\hat {\cal V}$. Hence, if the ground state is represented well by the reference SUHF at zeroth order, the eigenvalues of ${\bf A}$ are expected to always be positive except for those resulting from redundancies. 

\section{Computational details}
In this section, we describe computational details. Symmetry-projected calculations were performed with the {\sc Gellan} suite of programs,\cite{GELLAN} and SR  (MP2, CCSD, CCSD(T)) and CASPT2 calculations were carried out with {\sc Gaussian}\cite{g09} and {\sc Molpro},\cite{Molpro} respectively.  Since we deal with unrestricted determinants, i.e., eigenstates of $\hat S_z$, the integrations of $\alpha$ and $\gamma$ can be performed analytically.\cite{Jimenez12,Tsuchimochi16B} Hence,  all calculations presented used $N_g = 4$ grid points  only for the $\beta$ rotations,  which was found to be sufficient to obtain numerically exact $\bra \hat S^2\ket$. Spatial symmetry is ensured by performing one-shot symmetry projection. For triplet calculations, typically high-spin states are found to be slightly more favorable than low-spin states, although the difference is usually negligible. In some cases, they cannot represent the correct spatial symmetry, and low-spin states are therefore used. 

In EMP2 and SUPT2, we often employ the frozen core approximation, where core electrons are not correlated. This can be achieved by constrained SUHF (cSUHF),\cite{Tsuchimochi10B,Tsuchimochi11,Tsuchimochi16B} where natural orbitals with the largest occupation numbers are obtained as doubly-occupied closed-shell orbitals. To correctly specify the desired doubly-occupied orbitals in the energetical order, we then form the generalized Fock matrix and diagonalize only in this closed-shell space. Note that the generalized Brillouin theorem is no longer satisfied for these orbitals, so single excitations are included in the evaluation of the second-order energy. 

The linear equations of EMP2 and SUPT2 are solved with direct inversion of iterative subspace (DIIS).\cite{Pulay80,Pulay82} In each iteration, the computational complexity scales as ${\cal O}(N_g N_o^2N_v^3)$, where $N_o$ and $N_v$ are the numbers of occupied and virtual orbitals, respectively. Currently, we simply use diagonal elements for preconditioning, which is not an optimal choice. Therefore, the DIIS convergence is somewhat slow with the present implementation. Nevertheless, other preconditioning schemes are available to improve the convergence behavior,\cite{Tsuchimochi17B} and we will test and report their performances in a separate paper.

\section{Illustrative calculations}\label{sec:Results}
\subsection{Single bond dissociation: HF}
\begin{figure}
\includegraphics[width=80mm]{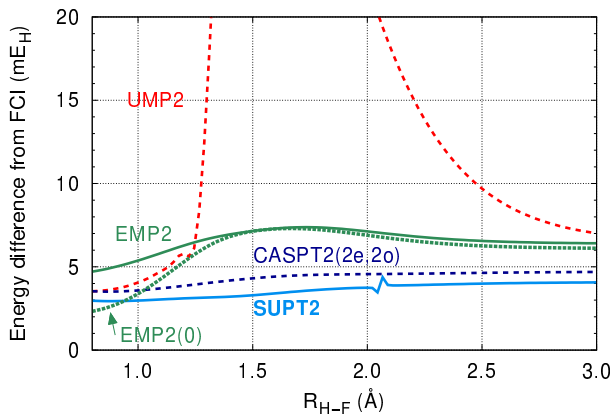}
\caption{Energy differences $E-E_{\rm FCI}$ in m$E_{\rm H}$ for several perturbation schemes in the HF potential energy curve computed with the 6-31G basis.} \label{fig:HF_dif}
\end{figure}

We use the HF molecule as our first test case. The 6-31G basis set is used,\cite{6-31G} and the F $1s$ orbital is frozen. Figure \ref{fig:HF_dif} shows the energy differences of several methods against FCI. As is well known, UMP2 gives a sharp derivative discontinuity at the Coulson-Fischer point, where a HF determinant breaks spin-symmetry. Passing this point, broken-symmetry UMP2 gives a substantial error and becomes completely unreliable. Interestingly, EMP2 and EMP2(0) are very similar in energy to each other, showing almost no improvement of the former. This similarity is also seen in many other cases, indicating that the broken-symmetry Fock matrix already well represents $\hat {\tilde {\mathcal H}}_0$ used in EMP2. Still, in general, EMP2(0) gains more correlation energy around the equilibrium bond length (c.a. 0.95 {\AA}), while both EMP2(0) and EMP2 tend to become less accurate when a molecule is stretched. Therefore, overall, the potential energy curve of EMP2 is more parallel to FCI. As a matter of fact, the non-parallelity-error (NPE), which is defined as the difference between the maximum and minimum errors from FCI, is 2.7 m$E_{\rm H}$ for EMP2 and 5.0 m$E_{\rm H}$ for EMP2(0). 
\begin{figure}
\includegraphics[width=80mm]{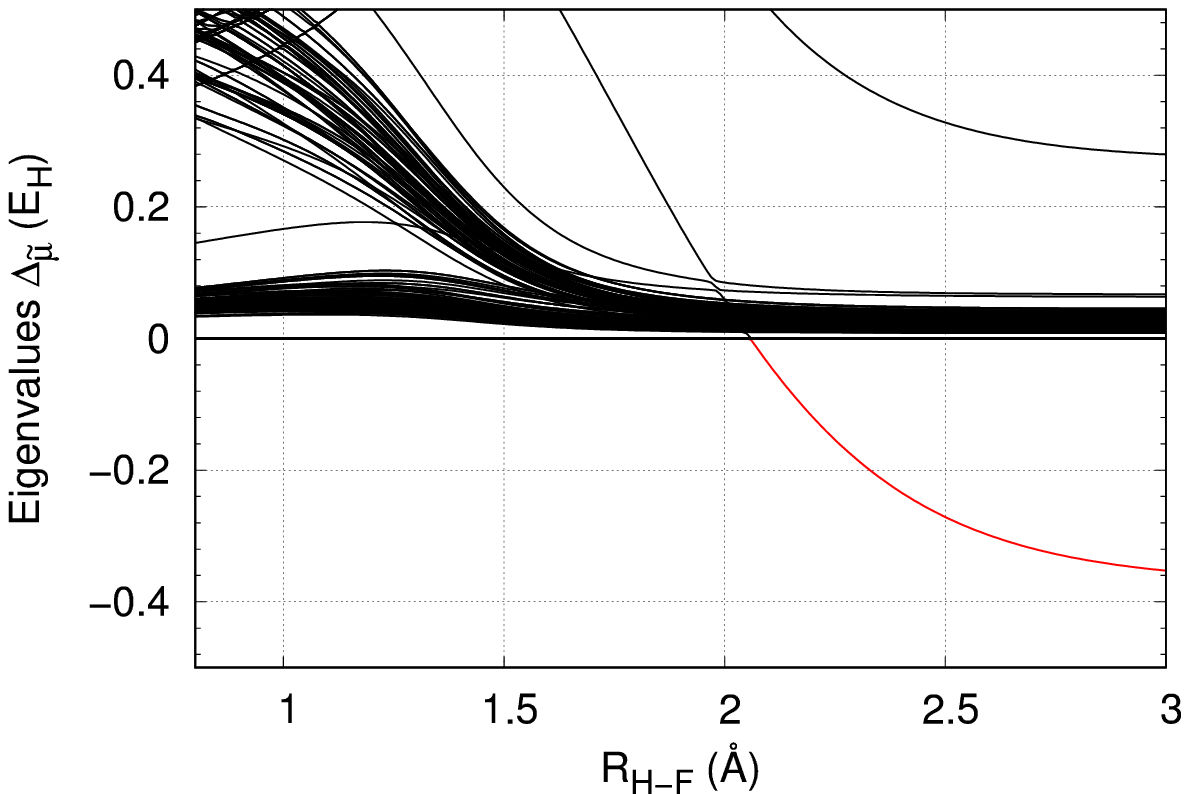}
\caption{Eigenvalues $\Delta_{\tilde \mu}$ of {\bf A} for HF in SUPT2.} \label{fig:HF_eig}
\end{figure}
While EMP2 and EMP2(0) both outperform SUHF, whose NPE is 13.8 m$E_{\rm H}$, their improvements are not impressive, given that CASPT2 with the minimal active space of ($2e,2o$) for single-bond breaking is even more accurate with an NPE of 1.2 m$E_{\rm H}$. Because CASSCF ($2e,2o$) is a subset of SUHF,\cite{Tsuchimochi11,Tsuchimochi16B} it is expected that a PT2 from SUHF is comparable to or better than CASPT2 ($2e,2o$). This is indeed the case for SUPT2, which gives less errors along the dissociation path. While the SUPT2 curve looks encouraging, it turns out to be discontinuous at approximately 2.05 {\AA}. To inspect the sudden change in energy, the eigenvalues $\Delta_{\tilde \mu}$ of {\bf A} are plotted in Figure \ref{fig:HF_eig}. As can clearly be seen, one of the eigenvalues becomes negative at the said point, responsible for the divergence in the second-order energy. It is noteworthy that a negative denominator ($\Delta_{\tilde\mu} < 0$) itself does {\it not} cause any problem, but an eigenvalue crossing zero is what is at stake.  The characteristic of this nearly zero eigenvalue is different from that of other {\it essential} zero eigenvalues, which are caused by redundancies and can be easily removed because the corresponding $V_{\tilde \mu}$ are also exactly zero in Eq.~(\ref{eq:Tmu}).

Since CASPT2 ($2e,2o$) does not show such a divergence for this simple  molecule, it is most likely that the intruder state in SUPT2 corresponds to fully-internal excitations (ones within the active space) in CASPT2. In this sense, the intruder state problem seems more severe in SUPT2 than in CASPT2 because we never distinguish excitation classes in the former. To remove this intruder state from SUPT2, either a real level shift of $\epsilon \approx 0.2$ $E_{\rm H}$ or an imaginary level shift was required; otherwise, the energy divergence persists. In passing, as mentioned above, both EMP2(0) and EMP2 do not suffer from intruder states.  While the performance of SUPT2 is relatively satisfactory when the amplitudes are stable, the intruder state problem is a significantly unfavorable feature. In the next section, we will investigate this problem in more detail and show that the imaginary shift scheme appears to be the best compromise.

\subsection{Multiple bond dissociation: H$_2$O and N$_2$}
In this section, we focus on the symmetric dissociation of H$_2$O and the triple-bond breaking of N$_2$ as more complicated cases. Again, we use the 6-31G basis set and freeze the 1s orbitals of O and N as in the previous section. 

\begin{figure}
\includegraphics[width=85mm]{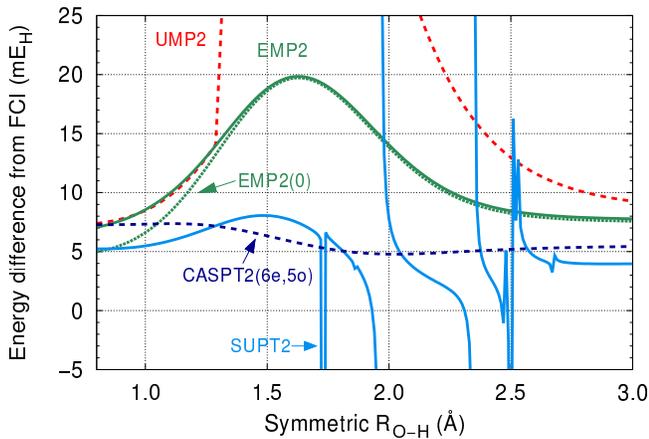}
\caption{Same as Figure \ref{fig:HF_dif} but for H$_2$O.} \label{fig:H2O_dif}
\end{figure}

\begin{figure}
\includegraphics[width=85mm]{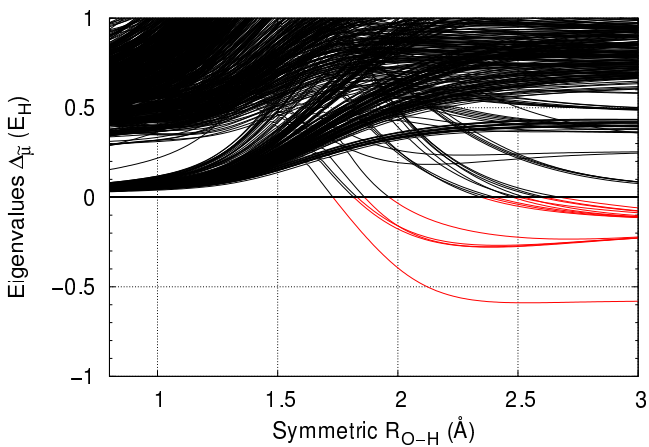}
\caption{Same as Figure \ref{fig:HF_eig} but for H$_2$O.} \label{fig:H2O_eig}
\end{figure}

In Figure \ref{fig:H2O_dif}, the energy error against FCI is plotted every 0.01 {\AA} from R$_{\rm O-H} = 0.8$ {\AA} to 3.0 {\AA} for the symmetric dissociation of H$_2$O. Most of the conclusions we drew in the previous section still hold here. The second-order energies computed with EMP2 and EMP2(0) are basically the same, while the latter is slightly larger at short bond lengths. Clearly, there are many more intruder states in SUPT2 compared to the case of the HF molecule, making its potential curve very unstable. Again, they can be understood as divergence in amplitudes.  To see this, the eigenvalue profile of ${\bf A}$ in SUPT2 for H$_2$O is plotted in Figure \ref{fig:H2O_eig}. Note that the discontinuous positions of SUPT2 in Figure \ref{fig:H2O_dif}  exactly correspond to the points where one of $\Delta_{\tilde \mu}$ crosses zero in Figure \ref{fig:H2O_eig}. 

At this point, a remedy is indispensable to obtain meaningful potential curves with SUPT2. We have tested real and imaginary level shifts with $\epsilon = 0.1, 0.2, 0.3,$ and $0.4$ $E_{\rm H}$ to alleviate the ill-behaved potential curve, and Figure \ref{fig:H2O_Imag} shows their energy differences from FCI where the level-shift corrected energy is evaluated according to Eq.~(\ref{eq:Hylleraas}). As expected, introducing a real level shift tends to quench the singularities as $\epsilon$ becomes larger, and it appears that $\epsilon=0.3$ is sufficient to obtain a smooth curve for the present case. The second-order  energy becomes slightly less accurate with $\epsilon$, but this happens to a similar extent at all bond distances. In Table \ref{tb:H2O_NPE}, we have tabulated the NPEs for the H$_2$O curves computed with the uncorrected and corrected second-order energies, $E_2$ and $\cal L$ (Eqs.~(\ref{eq:E2}) and (\ref{eq:Hylleraas})). The level-shift correction is essential to keep the qualitative results of rSUPT2. As such, we will report only the level-shift corrected energy $\cal L$ below, if not mentioned otherwise. However, for $\epsilon = 0.1$ and $0.2$, the use of $\cal L$ does not cure the intruder state problem at all, and the divergence behavior is often amplified because $\bra \psi_1|\psi_1\ket \gg 1$. Unfortunately, it is not possible to estimate  a value that removes all singularities in potential energy surfaces {\it a priori}; therefore, a trial and error approach is required.
 
In this regard, the imaginary shift scheme is more promising. It can be shown that, away from the singularities, the energy error induced by real $\epsilon$ is on the order of $\left(\frac{\epsilon}{\Delta_{\tilde \mu}}\right)$ for $\Delta_{\tilde\mu} \gg 1$, whereas that for imaginary $i\epsilon$ is $\left(\frac{\epsilon}{\Delta_{\tilde \mu}}\right)^4$.\cite{Forsberg97, Park19, footnote_SUPT2}  Furthermore, the imaginary level shift is singularity-free. All these features are illustrated 
by Figure \ref{fig:H2O_Imag}, where the results for $i\epsilon= 0.1i, 0.2i, 0.3i,$ and $0.4i$ are all continuous and smooth.
 The energy error does not grow  with an increase in $\epsilon$ as significantly as for the real shift. As a result, NPEs are all reasonable for different $i\epsilon$ with iSUPT2 (Table \ref{tb:H2O_NPE}). Still, as can be seen, $\epsilon$ should not be too small or large in the imaginary shift scheme, and the recommended value range is $i\epsilon = 0.3 i \sim 0.5 i$.

\begin{figure}
\includegraphics[width=85mm]{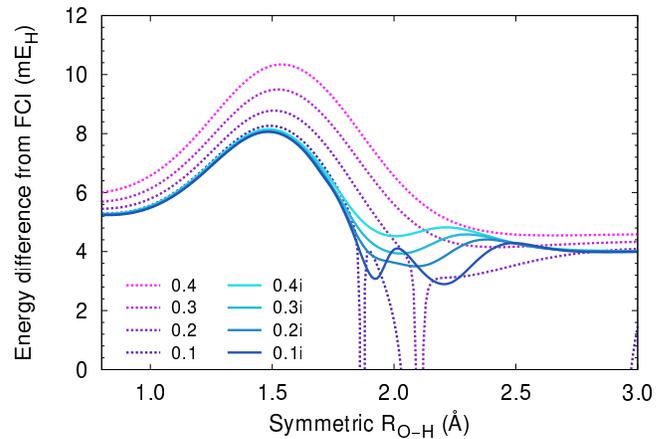}
\caption{Energy differences $E-E_{\rm FCI}$ in m$E_{\rm H}$ for several level-shift values in the symmetric dissociation of H$_2$O. Red curves are real shifts, whereas blue curves are imaginary shifts. The second-order energy is corrected with Eq.~(\ref{eq:Hylleraas}).} \label{fig:H2O_Imag} \label{fig:H2O_Imag}
\end{figure}

\begin{table}
\caption{NPEs of level-shifted SUPT2 for the symmetric dissociation of H$_2$O (m$E_{\rm H}$). }\label{tb:H2O_NPE}

\begin{threeparttable}[t]
\renewcommand{\arraystretch}{1.2}
\begin{tabular}{crrrrr}
\hline\hline
& \multicolumn{2}{c}{Uncorrected $E_2$} & &\multicolumn{2}{c}{Corrected $\cal L$} \\
\cline{2-3}\cline{5-6}
$\epsilon /E_{\rm H}$ &  Real  & Imag. & & Real & Imag.\;\;\;\\
 
\hline  
0.1 &---\tnote{a}  & 4.6 &&---\tnote{a}&  5.2\\
0.2 &---\tnote{a}   & 4.3& &---\tnote{a}&  4.6\\
0.3 & 9.7&4.6 & & 5.3& 4.2\\
0.4 & 11.1 & 5.0& & 5.9 & 4.1\\
0.5 &12.4&  5.6 & & 6.4 & 4.2 \\
0.6 &13.6& 6.1&& 7.0  &4.3 \\
\hline\hline
\end{tabular}
 {\footnotesize
\begin{tablenotes}
\item[a] Diverged.
\end{tablenotes}
}
\end{threeparttable}

\end{table} 

\begin{figure}
\includegraphics[width=85mm]{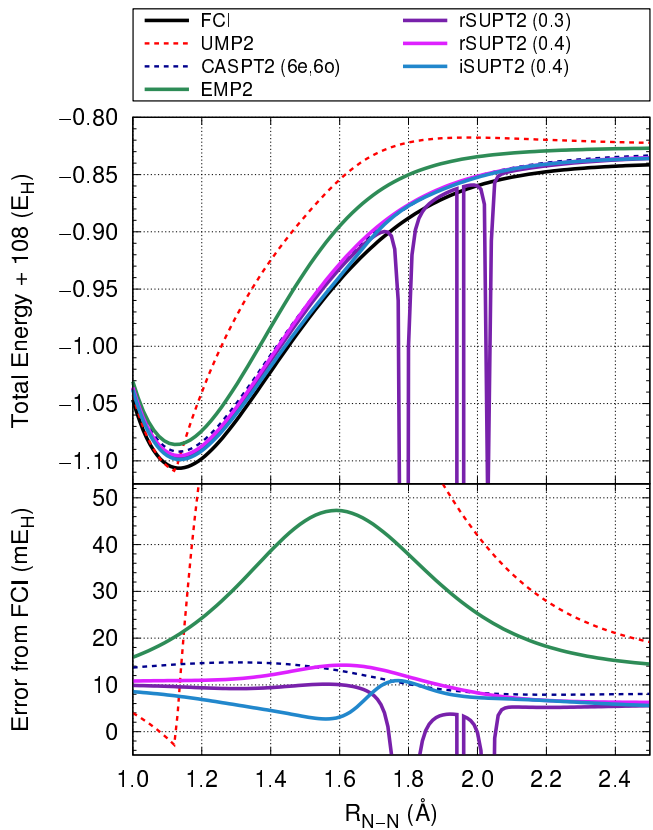}
\caption{{\it Upper panel}: potential energy curves of N$_2$ computed with several methods using the 6-31G basis. {\it Lower panel}:  Energy error from FCI.} \label{fig:N2}
\end{figure}

Now, we turn our attentions to N$_2$. This molecule is more challenging than HF and H$_2$O, and has been used to benchmark several MR methods.\cite{Forsberg97,Krogh01,Yanai06,Hanauer11,Manni14,Tsuchimochi16A} The upper panel of Figure \ref{fig:N2} shows the potential energy curves computed by FCI and different PT2 schemes, where we have used an active space of ($6e,6o$) for CASPT2, and employed $\epsilon = 0.3, 0.4$ and $i\epsilon =0.4i$ for the level-shift  in SUPT2.  We omit EMP2(0) because its energy is almost identical to that of EMP2. For the real-shifted SUPT2,  $\epsilon = 0.4$ is needed to remove all singularities. Therefore, with a real shift of $\epsilon = 0.3$, SUPT2 produces pronounced peaks. Once an appropriate value is used to eliminate singularities, real- and imaginary-shifted SUPT2 are similar in performance, and their potential energy curves are almost indistinguishable from each other. In the lower panel of Figure \ref{fig:N2}, we have plotted the energy differences from FCI for N$_2$. For $\epsilon = 0.4$ and $i\epsilon = 0.4 i$, the errors of SUPT2 are considerably smaller than those of EMP2. The correlation energies obtained with these level shifts are akin to those of CASPT2, giving a satisfactory description of triple-bond breaking.

\begin{table*}
\caption{Comparison of several methods for NPEs of HF, H$_2$O, and N$_2$ with respect to FCI (m$E_{\rm H}$).}\label{tb:NPE}

\begin{threeparttable}[t]
\renewcommand{\arraystretch}{1.2}
\begin{tabular}{crrrrrrrrr}
\hline\hline
&UMP2 & CASPT2\tnote{a} & SUHF & EMP2(0) & EMP2 & rSUPT2\tnote{b} & iSUPT2\tnote{c} & ECISD \\
\hline
HF & 39.3 & 1.2 & 13.8 & 5.0 & 2.7 & 1.1 & 1.1 & 0.9 \\
H$_2$O & 66.2 & 2.6 & 67.9 & 14.7 & 12.8 & 5.3 & 4.1  & 3.8\\
N$_2$ & 100.2 & 6.9 & 104.1 & 38.0 & 35.1& 8.0 & 8.2 & 15.6\\
\hline\hline  
\end{tabular}
 {\footnotesize
\begin{tablenotes}
\item[a] Active space: $(2e,2o)$ for HF, $(6e,5o)$ for H$_2$O, and $(6e,6o)$ for N$_2$.
\item[b] Level-shift value: 0.2, 0.3, and 0.4 for HF, H$_2$O, and N$_2$, respectively. 
\item[c] Level-shift value: $0.4 i$ for all molecules.
\end{tablenotes}
}
\end{threeparttable}

\end{table*}

Finally, we close this section by summarizing the NPEs of HF, H$_2$O, and N$_2$ for each method with 6-31G in Table \ref{tb:NPE}.  From the table, the remarkable strength of SUPT2 should be clear; although it requires a proper treatment of singularities, the level-shifted SUPT2 rivals CASPT2 in accuracy. For N$_2$, SUPT2 even outperforms ECISD, at only a fractional computational cost, indicating its potential. In particular, iSUPT2 is more advantageous than rSUPT2 in that it is capable of removing all singularities independent of $\epsilon$.

\subsection{Spectroscopic constants of N$_2$}\label{sec:N2}
While we have seen that both EMP2 and SUPT2 can treat both static and dynamical correlation effects reasonably well and can describe molecular dissociations, it is also important for them to be able to predict molecular properties, such as spectroscopic constants. For this purpose, we continue to use the N$_2$ molecule as the test system. We employed the aug-cc-pVQZ basis\cite{augcc} to compute the equilibrium bond length $R_e$, vibrational frequency $\omega_e$, and dissociation energy $D_e$, and compared the results with experiments.\cite{CRC} Although $D_e$ is calculated by the super-molecular approach, i.e., $D_e=E[100 \mbox{\AA}] - E[R_e]$, the size-consistency errors ($E[100  \mbox{\AA}] - 2 E[{\rm atom}]$  where quartet spin-projection is performed for atoms) are less than 0.02 kcal/mol for all methods.  The almost negligible size-consistency errors might come as a surprise, but are attributed to the character of the underlying broken-symmetry UHF determinant $\Phi_0\ket$ at the dissociation limit, which is a mixture of singlet, triplet, quintet, and septet, all nearly degenerate in energy. Since UHF is known to be size-consistent for the N$_2$ dissociation into two quartet atoms, which have almost no spin-contamination, the singlet SUHF energy is naturally very close to the sum of the septet spin-projected atoms. This is how SUHF breaks valence bonds in general.

As shown in Table \ref{tb:N2_spec}, as expected, CCSD(T) is most accurate and achieves ``chemical accuracy'' for all constants.\cite{Purvis82} While MP2 shows improvements over HF, it turns out that it overestimates the correlation energy ($E[R_e] = -109.39369$ $E_{\rm H}$), especially when compared to CCSD ($E[R_e] = -109.38684$ $E_{\rm H}$). Consequently, the equilibrium bond length and dissociation energy are also overestimated: by $0.013$ {\AA} and $8.1$ kcal/mol, respectively. The vibrational frequency $\omega_e$ is largely underestimated by 137 cm$^{-1}$. From these results, it is concluded that the MP2 level of theory is insufficient to describe the equilibrium of N$_2$. 

It is found that both SUHF and CASSCF ($6e,6o$) yield results far better than those of HF, indicating that it is quite advantageous to treat N$_2$ with a multi-determinant wave function, even at equilibrium. SUHF is still less accurate than CASSCF ($6e,6o$) because it lacks some dynamical correlation within the incomplete active space. This fact is directly reflected in their energy difference, which is more than 60 m$E_{\rm H}$. However, SUPT2's ability to treat fully-internal excitations means that it is able to capture the missing dynamical correlation at zeroth order; with a level-shift of $0.4i$, SUPT2 delivers a total energy very similar to that of CASPT2. The computed spectroscopic constants are in excellent agreement between these methods. They also resemble CCSD, although $\omega_e$ predicted by CCSD is inferior to those by iSUPT2 and CASPT2 ($6e,6o$). We find that rSUPT2 with $\epsilon=0.25$ also gives almost the same results as these methods, including in the total energy; however, its potential curve contains a few singularities, rendering its applicability somewhat questionable. 

EMP2(0) and EMP2 produce less correlation energies at equilibrium than  SUPT2 and CASPT2, by c.a. 10 m$E_{\rm H}$; however, at the dissociation limit, their energies are even more underestimated, and the computed $D_e$’s therefore happen to be in better agreement with the experimental value.  Nonetheless, it is clear that their descriptions are not satisfactory for $R_e$, which show almost no improvement over SUHF. The computed $\omega_e$ are even worse than that of SUHF. Overall, SUPT2 with an appropriate level shift prevails over EMP2(0) and EMP2 in predicting the spectroscopic constants of N$_2$.

\begin{table*}
\caption{Spectroscopic constants of N$_2$ computed with the aug-cc-pVQZ basis set.}\label{tb:N2_spec}
\begin{threeparttable}[t]
\renewcommand{\arraystretch}{1.2}
\begin{tabular}{lrrrrr}
\hline\hline
Method & $R_e$/{\AA} & $\omega_e$/cm$^{-1}$ & $D_e$/kcal mol$^{-1}$ & $E[R_e]$/$E_{\rm H}$\\
\hline
HF & 1.066 & 2729 &  122.0 & -108.99493\\
MP2 & 1.111 & 2202 & 236.5 & -109.39369\\
CCSD & 1.093  &2434  & 214.4 & -109.38684\\
CCSD(T) & 1.100 & 2355 & 223.5 & -109.40724 \\
SUHF & 1.090 &2410 & 159.1 & -109.06489\\
EMP2(0) &1.090 &2471 & 223.4 & -109.37420\\
EMP2 & 1.092 & 2453 &222.4  & -109.37291\\
rSUPT2 ($0.25$) &1.102&  2330 &214.5 & -109.38428 \\
iSUPT2 ($0.4$) & 1.102 &2317  & 214.5 & -109.38589\\
CASSCF ($6e,6o$) & 1.102 & 2351& 205.4 &-109.12770\\
CASPT2 ($6e,6o$) & 1.101 & 2334 & 215.1 & -109.38520\\
Exp.  & 1.098 & 2359 & 228.4 & \\
\hline\hline
\end{tabular}
\end{threeparttable}
\end{table*}

\subsection{Singlet-triplet splitting energies}
Excitation energy is an important quantity. There are approaches to treat excited states based on the PHF framework, such as linear-response theory\cite{Tsuchimochi15A} and nonorthogonal CI.\cite{Jimenez13A} However, since our PT2 methods are currently formulated in a state-specific way, it is not straightforward to apply them to excited states. Having said that, it is relatively easy to calculate the lowest state of a given spin symmetry. 

Recently, Rivero et al. benchmarked singlet-triplet splitting energies with several PHF methods, including SUHF.\cite{Rivero13} They showed that while SUHF's results are reasonable, further improvements can be achieved by breaking and restoring a variety of other symmetries, such as $\hat S_z$. This means that a balanced treatment of static and dynamical correlation effects is important for predicting accurate singlet-triplet gaps. Hence, it is interesting to ask how much advantage our second-order perturbation theories bring about in computing this quantity. 
   \begin{table*}
\caption{Computed singlet-triplet gaps of small systems in kcal/mol.}\label{tb:ST}
\begin{threeparttable}[t]
\renewcommand{\arraystretch}{1.2}
\begin{tabular}{lrrrrrrrrrrr}
\hline\hline
Method & 			C&		O	&	Si	&	NH	&	OH$^+$&	O$_2$	&	NF&\hspace{3mm}&	ME	&MAE \\
\hline
CASPT2 ($2e,2o$)&	29.1&	45.5&	17.5&	36.5&	49.8&	15.4&	32.0&&	-1.3&	1.6\\
CASPT2 (FV)	&	29.1&	45.5&	17.4&	37.0&	50.2&	22.8&	32.2&&	-0.1&	0.6\\
SUHF			&	22.3&	38.7	&	8.3	&	32.7&	45.2&	26.0&	31.6		&&-4.3	&5.3	\\
EMP2(0) 		&	29.1	&	45.4&	17.2&	35.8&	49.3&	25.2&	34.4&&	0.2&	0.6\\
EMP2			&	29.1&	45.3&	17.3&	35.7&	49.1&	24.9&	33.9&&	0.1&	0.7\\
SUPT2 			&	29.7&	46.2&	17.8&	36.7&	50.5&	24.3&	34.1&&	0.6&	0.7\\
rSUPT2 ($0.25$)	&	29.6&	46.2&	17.5&	36.7&	50.5&	24.6&	34.2&&	0.6&	0.7\\
iSUPT2 ($0.4$)	&	29.7&	46.2&	17.6&	36.8&	50.5&	24.2&	34.2&&	0.6&	0.6\\
Exp.			&	29.1&	45.2&	17.3&	35.9&	50.5&	22.6&	34.3&&	&	\\
\hline\hline
\end{tabular}
\end{threeparttable}
\end{table*}

\subsubsection{Atoms and diatomic molecules}
We first compute the ST splitting energies of atoms (C, O, and Si) and diatomic molecules (NH, OH$^+$, O$_2$, and NF). We use the aug-cc-pVQZ basis and the experimental geometries for the molecules.\cite{Slipchenko02} All electrons are correlated in our calculations. For these atoms, the ground state is a triplet $^3P$ state, whereas the lowest singlet state is $^1D$. For the molecules, we compute the adiabatic excitation energies of $^3\Sigma \rightarrow ^1\Delta$. The biradical nature of these systems poses a challenge for SR methods because their singlet states are qualitatively represented by two determinants, meaning very demanding triple excitations are required for quantitative accuracy. Consequently, standard post-HF methods, such as MP2 and CCSD, significantly overestimate the ST gaps.\cite{Tsuchimochi14}

Table \ref{tb:ST} presents the calculated ST gaps together with the mean errors (MEs) and mean absolute errors (MAEs) against the experimental values.  We have used two active spaces for CASPT2: $(2e,2o)$ and full-valence (FV) spaces. The former is the minimum space required to treat (two-determinantal) biradical systems, and triplet states are simply a single determinant of restricted open-shell HF. From the table, it is immediately clear that this small active space is not sufficient for the ST gap of O$_2$; the predicted value is 15.4 kcal/mol, and the error against the experimental value (22.6 kcal/mol) is 7.2 kcal/mol. The rather large error is ascribed to the fact that both singlet and triplet states are overly correlated in this system with CASPT2 ($2e,2o$). This imbalance was not fixed by a level-shift; CASPT2 ($2e,2o$) with $\epsilon=0.25$ still gave an ST gap of 15.8 kcal/mol. On the other hand, CASPT2 with the full-valence active space ($12e,8o$) yields an excellent result of 22.8 kcal/mol. Overall, the MAE of FV-CASPT2 is 0.6 kcal/mol, whereas that of CASPT2 ($2e,2o$) is 1.6 kcal/mol. However, it is apparent that it might not always be feasible to employ a full-valence active space. 
It is important to select active orbitals that are physically relevant, but they depend on various factors such as geometry and chemical reactions. After all, it still remains difficult to construct an appropriate active space, although many authors have suggested practical ways to ease this task.\cite{ Pulay88,  Bofill89, Jensen88, Abrams04, Stein16,Sayfutyarova17} 

SUHF does not usually require an active space to be chosen (except for the specification of core orbitals) and is therefore more flexible in this sense. For these rather simple examples, we found all PT2 schemes based on SUHF delivered similarly accurate descriptions. The difference between EMP2(0) and EMP2 is almost negligible, as was seen in the previous sections, and both achieved accuracies similar to that of FV-CASPT2. The maximum errors were obtained for O$_2$, but they are less than those of CASPT2 ($2e,2o$): $+2.6$ and $+2.3$ kcal/mol for EMP2(0) and EMP2, respectively. The chief difference between SUHF and CASSCF ($2e,2o$) in this system is that the anti-bonding $\pi_g$ orbitals are fractionally occupied in the former. The natural occupation numbers of SUHF are 0.012 and 0.031 for the singlet and triplet, respectively, implying that there is some contribution to static correlation that the minimum active space was not able to capture in CASSCF ($2e,2o$). 

For the tested systems, the SUPT2 amplitudes are stable without a level shift, and we can thus investigate the accuracy that the original SUPT2 potentially has to offer. For comparison, we have carried out SUPT2 calculations with three different level-shift conditions: $\epsilon = 0, 0.25$ and $i\epsilon = 0.4i$. As can be seen from Table \ref{tb:ST}, SUPT2 without a level shift provides results as accurate as EMP2.  Evidently, the accuracy of SUPT2 is almost unchanged when a level shift is introduced. The energy deviation caused by a level shift occurs in a balanced manner between singlet and triplet states (less than a few m$E_{\rm H}$ in all cases) such that the influence to the calculated excitation energy is negligible. 

Overall, both EMP2 and SUPT2 can successfully predict the ST gaps for the systems tested here, while CASPT2 is also accurate if the active space is properly chosen. However, for more complicated systems, EMP2 and SUPT2 show different trends, as will be demonstrated below.

  \begin{table}
\caption{Lowest singlet-triplet excitation energies for transition metal complexes (eV).}\label{tb:TM}
\begin{threeparttable}[t]
\renewcommand{\arraystretch}{1.2}
\begin{tabular}{lcc}
\hline\hline
& Ferrocene  & $\left[{\rm Fe(NO)(CO)}_3\right]^-$ \\
&$^3E''_1$	&$^3A_1$ \\			\hline
HF & 0.02 &--- \\
MP2 & 1.98 &---\\
CASSCF\tnote{a} & 0.97\tnote{b},  1.91\tnote{c} &  2.27\tnote{d}, 1.76\tnote{e}, 2.44\tnote{f}\\
NEVPT2\tnote{a} & 1.88\tnote{b}, 2.09\tnote{c} &  2.63\tnote{d}, 3.40\tnote{e}, 2.43\tnote{f}\\
SUHF & 2.03 & 3.30\\
EMP2 & 1.47 & 1.25 \\
iSUPT2\tnote{g} & 1.59 & 2.63 \\
Reference & 1.74\tnote{h} & 2.32\tnote{i}\\
\hline\hline
\end{tabular}
 {\footnotesize
\begin{tablenotes}
\item[a] Taken from Ref.~[\onlinecite{Sayfutyarova17}].
\item[b] Active space of $(10e,7o)$.
\item[c] Active space of $(18e,15o)$.
\item[d] Active space of $(10e,8o)$.
\item[e] Active space of $(14e,9o)$.
\item[f] Active space of $(16e,14o)$.
\item[g] Imaginary shift of $0.4i$ $E_{\rm H}$. 
\item[h] Experimental value from Ref.~[\onlinecite{Armstrong67}].
\item[i] MRCI+Q with an active space of ($14e,9o$), consisting of Fe $d$ orbitals and the NO $\pi$ and $\pi^*$ orbitals. Ref.~[\onlinecite{Lin15}].
\end{tablenotes}
}
\end{threeparttable}
\end{table}

\subsubsection{Transition metal complexes}
Transition metal complexes are challenging not only for SR methods but also for MRPT2, as the results typically depend on the choice of active orbitals. Here, we report the results of our methods on Ferrocene Fe(C$_5$H$_5$)$_2$ and $\left[{\rm Fe(NO)(CO)}_3\right]^-$ and compared the ST gaps with strongly contracted $N$-electron valence state PT2 (NEVPT2).\cite{ Angeli01A, Angeli01B, Angeli02} The geometries were taken from Refs.~[\onlinecite{Harding08}] and [\onlinecite{Lin15}], respectively. We used the cc-pVTZ basis set and froze the $1s$ orbitals of C, N, and O and the $1s, 2s, 2p, 3s,$ and  $3p$ orbitals of Fe in the PT2 calculations. Relativistic effects were not accounted for in this study because it has been reported that they do not significantly affect the results.\cite{Sayfutyarova17}  

For Ferrocene, the singlet state is dominated by a single configuration of Fe $d^6$. The lowest triplet state is doubly degenerate $^3E''_1$, mainly characterized as the $d$-$d$ transitions from ($d_{xy}, d_{x^2-y^2} $) to ($d_{xz}, d_{yz}$).\cite{Ishimura02} Symmetry-breaking and restoration within SUHF results in a triplet state that is dominantly $^3E''_1$ but is slightly mixed with $E'_2$ spatial symmetry. We performed SUHF followed by spin-constrained SCF calculations, where we optimized SUHF orbitals such that the lowest 42 and 43 orbitals were doubly occupied in the singlet and triplet states, respectively. 

For the complex anion $\left[{\rm Fe(NO)(CO)}_3\right]^-$, both the singlet ground state and lowest triplet state ($^3A_1$) are strongly correlated. A previous study indicated that the strong electron correlation arises from two degenerate bonding and anti-bonding orbital pairs, mainly composed of Fe $d$ and NO $\pi^*$. \cite{Klein14} In particular, the state $^3A_1$ cannot be described by a single determinant in principle. We used a low-spin representation of SUHF to treat this triplet state. Constrained optimization was conducted to yield 37 doubly occupied orbitals. 

Table \ref{tb:TM} lists the ST gaps computed with various methods. As mentioned above, Ferrocene may be treated with SR methods reasonably.\cite{Ishimura02,Harding08} Indeed, although the predicted ST gap is much too small at mean-field HF level of theory (0.02 eV), the MP2 dynamical correlation brings a significant improvement, yielding 1.98 eV, which is in good agreement with the experimental value of 1.74 eV.\cite{Armstrong67} However, $\left[{\rm Fe(NO)(CO)}_3\right]^-$ requires a multi-reference treatment, and we were not able to obtain an ST gap with these methods. 

Several active spaces were tested for CASSCF and NEVPT2 in Ref.~[\onlinecite{Sayfutyarova17}], to which the reader is referred for more details about the active spaces used. As can be seen, the NEVPT2 results are mostly accurate, except for $\left[{\rm Fe(NO)(CO)}_3\right]^-$ with ($14e,9o$), which results in a gap of 3.40 eV (the reference value of  MRCI+Q is 2.32 eV). 
This indicates the importance of selecting appropriate active orbitals. 

For a mean-field theory, SUHF drastically improves the ST gap of Ferrocene upon HF. We found that SUHF gains some portion of dynamical correlation, especially in the singlet ground state, resulting in a good opening of ST gap (2.03 eV). However, as SUHF overestimates the gap by approximately 1 eV for the complex $\left[{\rm Fe(NO)(CO)}_3\right]^-$, a balanced description between static and dynamical correlations is necessary. The dynamical correlation effects of EMP2 and SUPT2 tend to close the gap of SUHF. This is in contrast with the correlation effect of MP2 and NEVPT2, both of which predict larger gaps than their zeroth-order treatments. For both systems,  EMP2 overcorrects the gap from SUHF, especially for $\left[{\rm Fe(NO)(CO)}_3\right]^-$, where the gap is underestimated by more than 1 eV. On the other hand, SUPT2 with an imaginary shift of $0.4i$ offers accurate gaps compared to both SUHF and EMP2. Its results are also comparable to those of the highly sophisticated NEVPT2 approach. Finally, we have not tested EMP2(0) and rSUPT2, but we expect their results to be similar to those of EMP2 and iSUPT2, respectively. 

\subsection{Chromium dimer}
Describing the electronic structure of Cr$_2$ is notoriously challenging not only because it requires a considerable amount of static correlation at equilibrium but also because dynamical correlation plays a significant role. For this reason, only by highly sophisticated methods can its potential energy curve be computed with qualitative accuracy.\cite{Andersson94,Roos95,Stoll96,Dachsel99,Muller09,Kurashige11,Coe14,Purwanto15,Vancoillie16,Guo16,Sokolov16} 
\begin{figure}
\includegraphics[width=85mm]{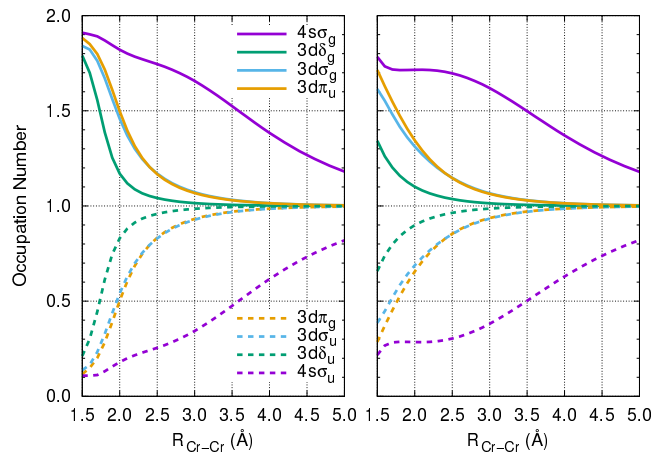}
\caption{Natural occupation numbers of Cr$_2$ with CASSCF ($12e,12o$) ({\it left}) and SUHF ({\it right}).} \label{fig:Cr2_NO}
\end{figure}
It is well known that the experimental potential energy curve of Cr$_2$ has a double-well structure\cite{Cr2Exp}; the first deep minimum corresponds to the $3d$-$3d$ bonding and the shallow, shelf-like region is ascribed to the dissociation of the $4s\sigma$ bond. Therefore, it is critical for a zeroth-order reference wave function to be capable of capturing these different bonding effects. 

Whether a method can describe such bonding effects is ensured by computing natural occupation numbers. The left and right panels of Figure \ref{fig:Cr2_NO} show the natural occupation numbers of CASSCF ($12e,12o$) and SUHF, respectively, computed with cc-pVQZ as a function of bond length. In both methods, the occupation numbers of the $4s\sigma_g$ and $\sigma_u$ orbitals slowly decay to one (which corresponds to bond dissociation), while those of the $3d$ bonding and anti-bonding orbitals show a rapid decay. Thus, SUHF gives a qualitatively correct description. Seemingly, the $3d$ occupation numbers of SUHF are more fractional (closer to one) than those of CASSCF at a short distance. This is attributed to the dynamical correlation effect captured within the CAS, which is mostly neglected in SUHF. For instance, the SUHF energy at $R=1.6$ {\AA} is higher than the CASSCF energy by 143 m$E_{\rm H}$, which is nevertheless reasonable given the N$_2$ case where SUHF misses a dynamical correlation energy of 60 m$E_{\rm H}$ (see Section \ref{sec:N2}). Importantly, it is expected that the fully-internal excitations in post-SUHF should exert their effectiveness for the missing dynamical correlation. Hence, with an appropriate post-SUHF scheme, one can expect to obtain a qualitatively correct potential energy curve of Cr$_2$.
\begin{figure}
\includegraphics[width=85mm]{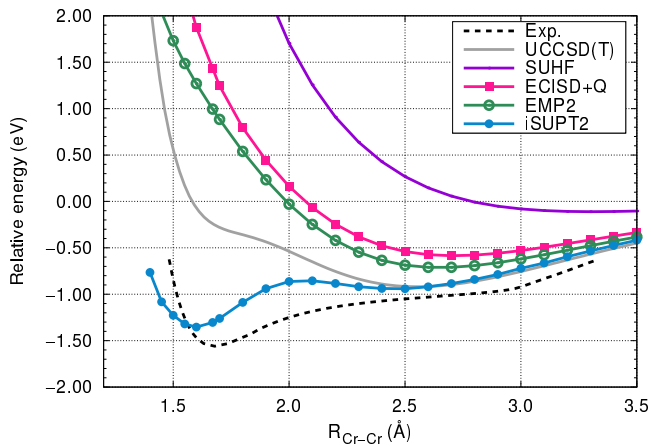}
\caption{Potential energy curves of Cr$_2$.} \label{fig:Cr2}
\end{figure}

In Figure \ref{fig:Cr2}, the potential energy curves of Cr$_2$ are plotted for several methods using the cc-pVQZ basis set. Here, $3p$ and $3d$ electrons are correlated, and no relativistic effect is taken into account. For spin-projection methods, we used 18 doubly-occupied orbitals. As expected, the SUHF curve is dissociative, meaning that a proper treatment of dynamical correlation is indispensable. The results of UCCSD(T), EMP2, and ECISD+Q are all disappointing, and they fail to predict the first minimum. On the other hand, it is intriguing that, unlike EMP2, the imaginary-shifted SUPT2 with $i\epsilon =0.4 i$ accounts for a large amount of dynamical correlation near the experimental equilibrium bond length $R_e = 1.68$ {\AA}, producing the global minimum. While the predicted bond distance is underestimated ($R_e = 1.60$ {\AA}), the double-well shape is well captured, and the computed dissociation energy of $D_e=1.36$ eV  is also comparable to the experimental estimate of $1.45-1.56$ eV.\cite{Cr2Exp1, Cr2Exp2, Cr2Exp3,Kurashige11} For a more detailed comparison, it is highly desirable to include the relativistic effect and to investigate the convergence in basis set size, which we plan to report in future work. 

Lastly, it is argued that CASPT2 ($12e,12o$) is not sufficient enough for Cr$_2$, and an active space of ($12e,28o$) is needed for a quantitative description.\cite{Kurashige11} The limitation of SUPT2 is that its zeroth-order reference SUHF is not systematically improvable unlike CASSCF, and our SUPT2 results therefore certainly cannot be made comparable to those of highly accurate CASPT2 ($12e,28o$). However, it is highly probable that  the use of spin-projected generalized HF (SGHF), in which further symmetry breaking and restoration of $\hat S_z$ is carried out,  will bring significant improvements over SUPT2, and it is thus interesting to pursue this direction in the future. In any case, the above results for Cr$_2$ clearly indicate the superiority of SUPT2 compared to EMP2 and CI.

\section{Discussions}\label{sec:Discussions}
That EMP2 becomes inferior for more strongly correlated systems is indicative that the excitations relevant to entangled (most symmetry-broken) orbitals are not treated as properly as in SUPT2. To investigate this implication, we have carried out the energy decomposition analysis for EMP2 and SUPT2, based on the double excitation class. To this end, we separate the SUHF natural orbital space into core ($c$), active ($a$), and virtual ($v$) spaces using appropriate occupation-number thresholds. Although the non-orthogonal nature of these excitations may not allow for the rigorous quantification of their contributions  because one cannot completely separate them in principle (especially if the Hylleraas functional is used to evaluate the energy), it is helpful to point out, even roughly, where the main difference between EMP2 and SUPT2 comes from.

 In cases where an SUHF wave function is a better {\it ansatz} than that of CASSCF, then SUPT2 is expected to offer more accurate results than CASPT2. We have already seen this for the HF molecule in Section 4.1 (CASSCF($2e,2o$) is a subset of SUHF). In this system, there are two active orbitals, and the energy contribution from the fully-internal double excitation, $(a, a) \rightarrow (a, a)$, is found to be negligible in both EMP2 and SUPT2 as expected. The total energy difference of $2\sim  4$ m$E_{\rm H}$ between the two methods is mainly attributed to the following two excitation classes; $(c, a) \rightarrow (v, a)$ and $(c, a) \rightarrow (v, v)$. For other excitations, either the energy contribution is virtually zero, or EMP2 and SUPT2 show almost identical energy contributions.

In general, the active space of SUHF is incomplete and is thought of as an approximation to CAS. Therefore, the fully-internal excitations in EMP2 and SUPT2 should play a vital role, perturbatively correcting the active space of SUHF. We argue that such a correction can be valid if the character of the SUHF active space is reasonably close to CAS. However, whether the correction is accurate or not also depends on the choice of the zeroth-order Hamiltonian. We found that, in most cases such as N$_2$, there is an appreciable difference between EMP2 and SUPT2 in the treatment of the fully-internal excitations. While these excitations capture a reasonable amount of correlation effects in SUPT2 and offer an improved approximation to CAS, they are not properly accounted for in EMP2. Although the contributions of other excitations such as $(c, a) \rightarrow (v, a)$ are also constantly underestimated in EMP2 compared to in SUPT2, they are relatively insignificant. The different treatments of the perturbative correction within the active space are the dominant contribution to the total energy difference, and it is concluded that the rather inferior behavior of EMP2 is attributed to the less accurate description of fully-internal excitations.

\section{Conclusions}\label{sec:Conclusions}
In this paper, we described second-order perturbation schemes with respect to spin-projected HF. The zeroth-order Hamiltonian of EMP2 was prepared as the Fock-like component of the projected Hamiltonian at each spin-rotation angle, whereas SUPT2 employed the generalized Fock operator constructed from the SUHF density matrix. The latter method almost always suffers from the intruder state problem, and we have discussed how one can remove singularities in practice by applying the level shift approach, especially with an imaginary shift value. These methods, together with  the  previously developed PT2, EMP2(0), were tested for several systems, including transition metal complexes. In general, the imaginary-shifted SUPT2 showed the best performance. It yielded potential curves that are reasonably parallel to those of FCI, and the computed singlet-triplet gaps were in good agreement with experimental values. We were also able to obtain a qualitative description of the Cr$_2$ molecule with SUPT2. On the other hand, the description of  the fully-internal space  in EMP2 is not satisfactorily accurate for difficult cases, such as multiple-bond dissociations and the spin-gap of $\left[{\rm Fe(NO)(CO)}_3\right]^-$. We therefore conclude that EMP2 is likely best for biradicaloid systems, and that SUPT2 stands as a preferable perturbative correction to SUHF.

With the good performance of SUPT2 demonstrated in this work, it is interesting to ask whether its accuracy still holds for the prediction of molecular properties. Our initial results for spectroscopic constants of N$_2$ are encouraging and support the validity of the SUPT2 method for such calculations. Computing molecular properties generally involves the relaxed density matrix and thus the derivatives of the total energy. Unfortunately, the level-shifted SUPT2 energy is not stationary with respect to the amplitudes. However, it is expected that the energy derivatives can be straightforwardly obtained by constructing an appropriate Lagrangian.\cite{Park19,Tsuchimochi17A} We are currently working on this task.

To achieve further quantitative accuracy, SUPT2 can be straightforwardly extended to SGPT2, second-order perturbation theory with SGHF. It has been shown that SGHF fixes many problems inherent in SUHF, and produces more accurate wave function and energy.\cite{Jimenez12,Rivero13} However, there are some additional complications that have to be addressed carefully, such as the treatment of the more general form of $\hat P$ in SGHF,\cite{Jimenez12} as well as the convergence of linear equation (\ref{eq:At=v}) with {\bf A} that is presumably dense in SGPT2. 

Other important developments include the generalization of our methods to excited states. In the present work, ground and excited states were treated separately in a state-specific manner. This clearly has a limitation in treating higher excited states and quasi-degenerate states, for which a multi-state formulation is required. Since SUHF uses a single (broken-symmetry) determinant, the single-particle picture is not completely lost. Indeed, we have exploited this fact when constructing the first-order wave function {\it ansatz} in this work. Hence, we are hopeful that it is not difficult to extend our schemes to excited states by combining with existing SR approaches, such as a second-order perturbative correction on CIS.\cite{HeadGordon94} 

\section*{Acknowledgment}
 This work was supported by JSPS KAKENHI  Grant Numbers JP17K14438 and JP18H03900, and MEXT as ``Priority Issue on Post-K computer (supercomputer Fugaku)'' (Development of new fundamental technologies for high-efficiency energy creation, conversion/storage and use).
We are also grateful for the computational resources of the K computer provided by the RIKEN Advanced Institute for Computational Science through the HPCI System Research project  (Project ID: hp160202, hp170259, hp180216, hp190175).
\bibliographystyle{aip}

\begin{thebibliography}{10}
\bibitem{footnote_SUPT2_2}
The linear dependence occurs because some of the linear combinations of
  broken-symmetry singles and doubles do not possess the designated
  spin-component, and hence they are null upon the spin-projection. Hence, such
  zeroes appear in the same fashion regardless of the operator.
\bibitem{footnote_SUPT2}
We should however note that this is only a rough estimate because we treat real
  and imaginary level shifts in a slightly different manner, see above.
\end{thebibliography}

\end{document}